# A Review of 1D Convolutional Neural Networks toward Unknown Substance Identification in Portable Raman Spectrometer

**M. Hamed Mozaffari and Li-Lin Tay**
*Metrology Research Centre, National Research Council Canada, Ottawa, ON K1A0R6, Canada*

**Abstract:** Raman spectroscopy is a powerful analytical tool with applications ranging from quality control to cutting edge biomedical research. One particular area which has seen tremendous advances in the past decade is the development of powerful handheld Raman spectrometers. They have been adopted widely by first responders and law enforcement agencies for the field analysis of unknown substances. Field detection and identification of unknown substances with Raman spectroscopy rely heavily on the spectral matching capability of the devices on hand. Conventional spectral matching algorithms (such as correlation, dot product, etc.) have been used in identifying unknown Raman spectrum by comparing the unknown to a large reference database of known spectra. This is typically achieved through brute-force summation of pixel-by-pixel differences between the reference and the unknown spectrum. Conventional algorithms have noticeable drawbacks. For example, they tend to work well with identifying pure compounds but less so for mixture compounds. For instance, limited reference spectra inaccessible databases with a large number of classes relative to the number of samples have been a setback for the widespread usage of Raman spectroscopy for field analysis applications.

State-of-the-art deep learning methods (specifically convolutional neural networks CNNs), as an alternative approach, presents a number of advantages over conventional spectral comparison algorism. With optimization, they are ideal to be deployed in handheld spectrometers for field detection of unknown substances. In this study, we present a comprehensive survey in the use of one-dimensional CNNs for Raman spectrum identification. Specifically, we highlight the use of this powerful deep learning technique for handheld Raman spectrometers taking into consideration the potential limit in power consumption and computation ability of handheld systems.

## 1 Introduction

Raman spectroscopy is a powerful analytical technique with applications ranges from planetary science to biomedical researches. For example, it has been employed in planetary exploration of extraterrestrial targets (Carey et al., 2015), identification of unknown substances in pharmaceutical, polymers, forensic (Yang et al., 2019), environmental (Acquarelli et al., 2017), and food science (Fang et al., 2018), and classification of bacteria, cells, biological materials (Fang et al., 2018; Kong et al., 2015). Recent advances, such as the development of microfabrication, faster computational resources have transformed the Raman spectroscopy from laboratory-based method to one that is being increasingly deployed in the field (Rodriguez et al., 2011). Lightweight portable (Jehlička et al., 2017) or handheld (Chandler et al., 2019) Raman spectrometers are applied for a wide range of applications including medical diagnostics for assisting physicians in identification of cancerous cells (Jermyn et al., 2017; Kong et al., 2015) and in the detection and identification of unknown substances such as explosives, environmental toxins and illicit chemicals (Weatherall et al., 2013). Raman spectroscopy has a number of advantages. It is reproducible with a finger-print spectrum unique to an individual molecule, and it is applicable to any optically accessible substances in various physical states with minimal sample preparation. However, miniaturized field-portable Raman spectrometers are often not as flexible as laboratory-grade grating Raman microscope. Design constraints imposed on the miniaturized optical components and power consumption considerations imposes operational limitations in the handheld spectrometers. Detection and identification of unknown substances with fieldable Raman spectrometers has a number of challenges: the requirement of reference databases with limited available memory, power consumption considerations for sustained field analysis, and limited computational capability results in slow detection time. To make matters more complicated, unknown substances encountered in the field are often mixtures, which presents significant challenge in the conventional spectral identification algorithms.

Portable Raman spectrometers tend to have lower spectral resolution compared to large focal length spectrometer commonly seen in the micro-Raman systems. Therefore, some critical spectral signature of target materials might not be resolved in data acquired by the lower resolution handheld spectrometers. With improvements in laser, optical components, gratings and detectors, handheld spectrometers continue to improve its performances. However, hardware improvements sometimes come with the compromise in size, affordability and power budget of the unit. On the other hand, improvement in the software has not being a high priority until recent years. This is in part driven by the rapid development of new computational tools in data analytics. Deep learning algorithms have been becoming a popular technique in solving pattern recognition in image processing and computer vision applications. Specifically, Convolutional Neural Networks (CNNs) have demonstrated that DL can discover intricate patterns in high dimensional data, reducing the need for manual effort in preprocessing and feature engineering (Lecun et al., 2015).

Advanced, automatic, and real-time CNN models are promising alternatives for improving the performance of spectral matching in handheld Raman spectrometers. Previous researches revealed that a deep learning model such as CNN has the capability of extracting meaningful information from raw low-resolution spectra without any preprocessing of data. In this article, we survey the recent publications that use a One-dimensional (1D) CNN model for the analysis of Raman spectral data. We investigate a few examples of deep learning models developed specifically for the handheld Raman spectrometers covering different applications. Here, we will be focused on the 1D CNN models that are readily applicable in the fieldable Raman spectrometers. Application of DL for more generalized Raman Spectroscopy and Raman imaging can be found in the two comprehensive reviews by (Lussier et al., 2020; Yang et al., 2019). This review is organized as follows. In section 2, we will introduce the concept of DL with a focus on 1D CNN in section 2.3. Section 3 discusses spectral matching and classification in detail. It starts with a review of conventional algorithms followed by applying 1D CNN in spectral matching. Section 4 concludes the study with our remarks on future potential of 1D CNN in field portable spectrometers.

## 2 Artificial Intelligence and Deep Learning

### 2.1 Machine Learning Era

The goal of artificial intelligence (AI) is to train computers, tools, and robots to repeat a task similar or even better than human performance. Deep learning is one of the approaches to achieve AI where it can successfully tackle a multitude of complicated problems that otherwise could not be solved efficiently. Deep learning pervades in our daily life from web search techniques, commercial recommendations in e-commerce websites, diagnosis of diseases in medical science, natural language processing that enables computers to communicate with, predicting potential drug candidates. In general, it is the foundation in pattern recognition and data mining.

The two main common characteristics of DL methods are multiple layers of nonlinear architectures and more feature abstraction in successively higher layers (Deng, 2014). Deep learning popularity owes to advances in sensors and data digitization technologies, which enable scientists to access big databases for training. Moreover, the development of big data analysis techniques helps to solve the over-fitting problem of neural networks partially. Other reasons are recent developments of efficient optimization algorithms, advances in graphical or tensor processing units (GPUs or TPUs) (Jouppi et al., 2017), cost-effective cloud computing infrastructure, the emergence of popular deep learning competitions such as ImageNet and Kaggle, and using parameters of pre-trained models instead of training a pure neural network from scratch using randomized initialization.

Machine learning (ML) techniques, designed before the deep learning era, have shallow-structure architectures, which usually consist of one or more layers of nonlinear feature transformations. Examples of those shallow architectures can be enumerated as Gaussian mixture models (GMMs) (Reynolds et al., 2000), linear or nonlinear dynamical systems (Svensson & Schön, 2017), conditional random fields (CRFs) (Zheng et al., 2015), maximum entropy models (Och & Ney, 2001), support vector machines (SVMs) (Drucker et al., 1999), logistic regression (Stoltzfus, 2011), and multi-layer perceptron (MLP) (Gori & Scarselli, 1998) with a single hidden layer including extreme learning machines (ELMs) (Deng, 2014).

### 2.2 Deep Learning Revolution

Although traditional ML methods have been used successfully and efficiently to address many pattern recognition problems, significant improvements are necessary to tackle complex real-world applications. Conventional techniques cannot adequately solve image processing problems such as image segmentation, image registration, classification of big data, real-time object recognition, and tracking in video frames. The ability of the human to solve these complicated problems prompted researchers to search for a better solution that simulates different models mimicking the pattern recognition and perception mechanism of the human brain.

Deep learning is one of the most active research fields in recent years. The objective is to learn features from the input dataset during the training stage and to predict instances in the testing stage using that learned knowledge. Deep hierarchical models with many layers (each of which followed by a nonlinear function) have shown better performance and accuracy than previous shallow-structured neural network models. An example of a successful simulation of the human brain is the use of multiple hidden layers with non-linearity in each layer (known as deep neural networks (DNNs)) in the traditional artificial neural networks (Deng, 2014; W. Liu et al., 2017). DNN is a multi-layer neural network with many fully connected hidden layers, which is usually initialized by unsupervised or a supervised pre-trained network. Various empirical studies have shown that using parameters of a pre-trained model instead of random initialization results in significantly better outcomes without the backpropagation and optimization difficulties. In general, using more hidden layers with many neurons and utilizing pre-trained models for the initialization of a deep neural network reduces the chance of trapping in poor local optima. Other factors can assist deep learning models in finding a better solution, including designing networks with efficient non-linearity like Rectified Linear Units (original or leaky)(Xu et al., 2015) and utilizing better optimization algorithm such as Stochastic Gradient Descent (SGD) (Ruder, 2016) and Adam optimization method (Kingma & Ba, 2014).

In a typical taxonomy, deep learning architectures might be classified into four classes: I) supervised learning methods (labelled and unlabelled training dataset are provided), II) unsupervised learning techniques (only unlabelled training dataset is available), III) semi-supervised deep networks as a combination of supervised and unsupervised approaches (Usually model is trained in two different steps), and IV) Reinforcement learning (training approach is based on action and rewards). There are lots of deep architectures in the context of deep learning, and each of which has been exploited for varieties of applications. In this review, we focus on the one-dimensional convolutional neural networks (1D CNNs) and its application for Raman spectroscopy classification task. 1D CNN is ideal for spectroscopy data where 1D spectrum contains local features such as sharp peaks. A comprehensive review of deep learning models can be found in reference (Lecun et al., 2015).

## 2.3  One-dimensional Convolutional Neural Networks

Convolutional Neural Networks was inspired by the functionality of the visual cortex in the animal brain, where it can automatically extract hierarchies of features in digitalized data such as signal or image from low-level to high-level patterns (Lecun et al., 2015). Mathematically, convolution is an operation that filters out the input data to indicate places of similarity between the filter and the input data. Similar to the brain structure, CNNs are artificial neural networks with alternating convolutional and sampling layers. For the case of an image, a two-dimensional convolution can be considered as applying and sliding a squared filter over the input image. For instance, if we are convoluting a filter of size $3 \times 3$ with an image, by setting all filter values to 1/9, the output results of the operation will be a moving average with a sliding window of length 3. In a CNN network model, these filter values are substituted by learnable parameters (also known as network weights) called "kernel". The kernel values are determined during the training of the network depend on the input data features. CNNs can be utilized in any dimension size, but usually, dimensions up to three are common in different applications, for example, 1D CNN for signals and time series (Ismail Fawaz et al., 2019), 2D for images (pixel-level) and matrices (Krizhevsky et al., 2017), and 3D for volumetric data (voxel-level) such as medical imaging databases (Krizhevsky et al., 2017).

A typical CNN model can extract features from input data with various levels of abstraction. As a deep neural network, the main component of a hierarchical layered CNN is convolutional blocks, which can extract different types of pattern in the input using various filters, followed by pooling layers which extract essential features from previous layers. This way of extracting features by different abstractions and sharing information from layer to layer creates a receptive field that helps the CNN model to be almost invariant to spatial transformations with a cheaper computational cost. For classification tasks, this way of network arrangement called encoder block, and output of encoders are extracted features in the input data such as features of the Raman spectrum. For this reason, to convert encoders into a classifier, it is common to append fully connected layers to the last layer of the network, followed by nonlinear functions such as SoftMax or Sigmoid. To avoid overfitting, batch normalization and dropout layers are added between layers. During the training stage of a CNN model, kernels of the network are first initialized randomly or different initialization techniques. This process of initialization can also be accomplished by using weights from a pre-trained model trained in advanced on a different dataset called transfer learning. Initialized kernels should be optimized using an optimization algorithm such as Adam or SGD utilizing the backpropagation technique.

1D CNN is a promising method for data mining of 1D signals where a limited number of data available and the whole signal information should be considered as the input data. 1D CNNs have recently been considered in limited applications such as personalized medical data classification and fault detection and identification in power electronics (Kiranyaz et al., 2019). The main advantage of this technique over previous artificial Neural Networks is that 1D CNN extracts features of a signal by considering local information instead of the whole signal in each network layer. This results in faster training of the network with a smaller number of trainable parameters, which cause less computational cost and power.

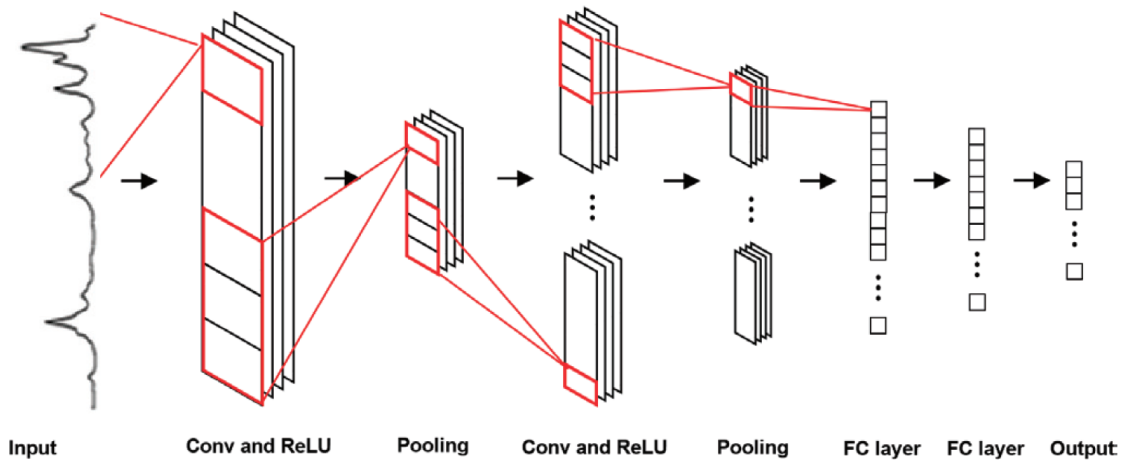

Figure 1 illustrates a typical 1D CNN models with consecutive 1D convolutional, down-sampling, and non-linearity layers.

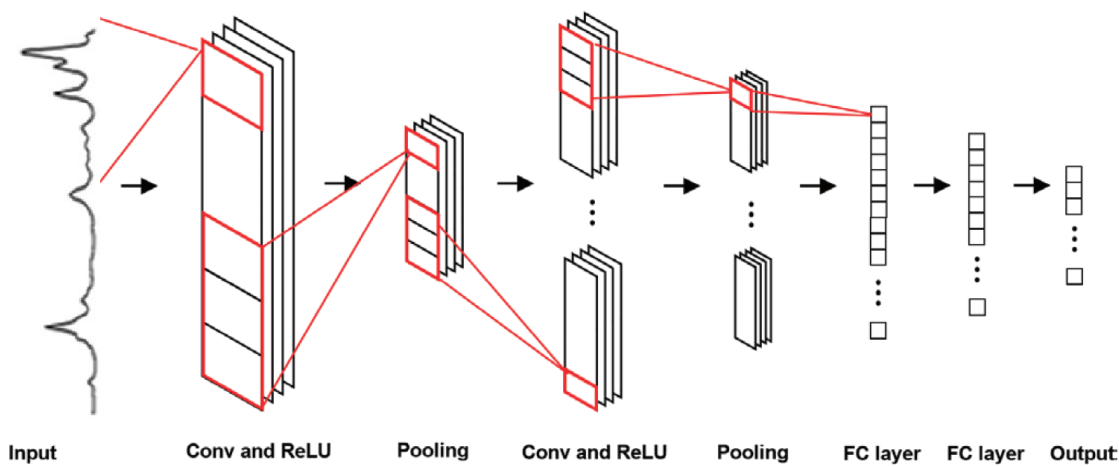

**Figure 1. A typical configuration of 1D CNN for spectroscopy data analysis.**

## 3 Spectral Matching and Classification

Spectral matching and identification is an enabling tool for modern spectroscopy. Several spectral matching algorithms have been developed since the 1970s (Grotch, 1970; Knock et al., 1970). Many of these early algorithms were developed for mass spectroscopy. Their applications soon expanded to cover vibrational spectroscopy. These conventional spectral matching approaches are iterative techniques. They are based on identifying the most similarities between the unknown and the reference spectra. In this section, we will briefly discuss the conventional spectral matching techniques followed by the use of 1D CNN in spectral matching for Raman spectroscopy.

### 3.1 Common Spectral Matching Techniques

From the pattern recognition point of view, methods applied for Raman spectral recognition can be classified into supervised and unsupervised techniques. The usual approach for unsupervised methods is to first transform data into a multi-dimensional space called embedding. Then reduce the number of dimensions using techniques such as Principle Component Analysis (PCA). Then, using an iterative method, such as Kth Nearest Neighbors (KNN), spectral data are clustered. The ultimate goal of a clustering method is to search the problem space and assign a class label for each sample. Therefore, the outcome is that each sample will have a maximum similarity within a class and a minimum similarity with members of other clusters. The paired comparison between samples is accomplished using a similarity criterion. **Error! Reference source not found.** outlines several examples of similarity measuring techniques. The performance of preprocessing has a crucial impact on the output results of similarity techniques. Similarity-based methods are divided into two classes of peak-feature, where a specific number of maximum peaks are compared, and full-spectrum matching with using all features (Sevetlidis & Pavlidis, 2019).

Various search algorithms have been developed to measure the similarity of measured Raman spectrum with spectra in reference databases such as correlation, Euclidean distance, absolute value correlation, and least squares (Kwiatkowski et al., 2010). These algorithms perform well in identifying pure components but fall short in the identification of mixture compounds. Multi-component substances require a more complicated database comprises of mixtures of pure components with different ratios (Fan et al., 2019). Raman spectrum often requires baseline correction prior to the application of spectral matching. There are a variety of methods for baseline correction, such as the Least-squares polynomial curve fitting for the subtraction of the baseline (Lieber & Mahadevan-Jansen, 2003). It is not uncommon that the spectrum is first smoothed prior to baseline correction. Comprehensive details of baseline correction and smoothing methods can be found in studies by (Y. Liu & Yu, 2016; Schulze et al., 2005) and (Eilers, 2003), respectively. **Error! Reference source not found.** outlines the few common methods used in baseline correction.

Table 1. List of famous similarity metrics for Raman spectral matching with reference databases.

| Name of Similarity Metric | References |
|---|---|
| Correlation Search and Cosine Similarity | (Carey et al., 2015; Howari, 2003; Kwiatkowski et al., 2010; Park et al., 2017; Stein & Scott, 1994) |
| Receiver Operating Characteristic (ROC) | (Fan et al., 2019) |
| Normalized Mean Square Error (NMSE) | (Malek et al., 2018) |
| High-Quality Index (HQI) | (S. Lee et al., 2013; Park et al., 2017; Rodriguez et al., 2011) |
| Absolute Difference Value Search (ADV) | (Kwiatkowski et al., 2010) |
| First Derivative Absolute Value Search (FDAV) | (Kwiatkowski et al., 2010) |
| Least Square Search (LS) | (Kwiatkowski et al., 2010) |
| First Derivative Least Square Search (FDLS) | (Kwiatkowski et al., 2010) |
| Euclidean Distance Search (ED) | (Howari, 2003; Kwiatkowski et al., 2010) |
| Correlation Coefficient (CC) | (Kwiatkowski et al., 2010) |
| Probability-based matching (PBM) | (Stein & Scott, 1994) |
| Composite Technique | (Stein & Scott, 1994) |
| Hertz method | (Stein & Scott, 1994) |
| Maximum likelihood estimator (MLE) | (Levina et al., 2007) |
| Hybrid Methods | (Khan & Madden, 2012) |

Table 2. List of popular techniques for baseline correction for Raman spectroscopy.

| Baseline correction methods | References |
|---|---|
| Polynomial Baseline Modeling | (Lieber & Mahadevan-Jansen, 2003; Juntao Liu et al., 2015) |
| Simulated-based methods (e.g., Rolling ball, Rubberband) | (Kneen & Annegarn, 1996) |
| Least Squares Curve Fitting | (Baek et al., 2015; Lieber & Mahadevan-Jansen, 2003; Z. M. Zhang et al., 2010) |

In unsupervised techniques, during the testing stage, an unknown sample can be recognized using a comparison between the target and reference spectra using the similarity metric used for clustering that database. Contrary, supervised techniques require ground truth labels (annotated manually by expert sample by sample) during training and testing stages. In the training stage, a supervised technique attempts to extract valuable features from the training dataset. For this reason, ground truth labels should be accurately provided as a training trend criterion. Then, a target spectrum can be easily classified in the testing stage using the knowledge acquired in the training phase. Current one-by-one matching techniques immensely depend on database quality and matching software performance (Carey et al., 2015). There is no such comprehensive database that includes all spectra, recorded with a similar standard. On the other hand, the performance speed of matching software will also drop when the reference database is extensive. This performance varies also depends on using full-spectrum, peak features, and preprocessing stages.

Principle machine learning (ML) techniques (see **Error! Reference source not found.** for a list of sample publications using ML) such as weighted KNN were tested for mineral recognition classification in a study by (Carey et al., 2015). Quantitative results of using KNN for Raman spectroscopy can be found in a study by (Madden & Ryder, 2003). In their research, a simple concept of ensemble learning using Artificial Neural Network was used for better prediction results. The advanced machine learning method which outperformed previous techniques of Raman classification was support vector machines (SVM) (Huang et al., 2017). Similar to other classifiers like logistic regression, SVM creates a hyperplane (decision boundary) that divides data into different classes. The hyperplane should optimally separate samples of different classes with a maximal margin between samples and the hyperplane. SVM is a powerful tool for classification tasks with limited classes, however, for problems with thousands of classes, training of a nonlinear version of SVM in practice is not possible (Jinchao Liu et al., 2017).

Before the widespread usage of Deep Learning methods for data classification, unsupervised methods such as random forest (RF) had been utilized extensively in many machine learning applications for high dimensional data. In a study by (Jinchao Liu et al., 2017), a fully connected version of ANN is considered as a weak method for spectral classification problems due to its shallow network. Two main issues of ANNs models with all satisfying results are over-fitting on training data and non-interpretability of ANN architecture, treated as a 'black box' (Acquarelli et al., 2017). In general, the main drawbacks of many previous methods for Raman spectral classification with a large number of classes are manual tuning during training and testing, and preprocessing and feature engineering, as well as over-fitting problem. A list of machine learning techniques used for spectroscopy can be found in a study by (Shashilov & Lednev, 2010).

Table 3. List of Machine Learning techniques used for Raman Spectroscopy. Note that references might not be the original research of each technique.

| Conventional Methods | References |
|---|---|
| Logistic Regression | (Nijssen et al., 2002) |
| K Nearest Neighbor (KNN) | (Madden & Ryder, 2003) |
| Weighted K Nearest Neighbor | (Carey et al., 2015) |
| Random Forest (RF) | (Sevetlidis & Pavlidis, 2019) |
| Artificial Neural Network (ANN) | (Acquarelli et al., 2017; Carey et al., 2015; Fan et al., 2019; Jinchao Liu et al., 2017) |
| Partial least square regression (PLSR) | (Arrobas et al., 2015; Wold et al., 2001) |
| Multiple linear regression (LR) | (Galvão et al., 2013) |
| Support vector machines for regression | (H. Li et al., 2009; Porro et al., 2008) |
| Gaussian process regression (GPR) | (T. Chen et al., 2007) |
| PLSR Linear Discriminant (PLS-LDA) | (S. Li et al., 2012) |
| Principle Component Analysis (PCA) | (Vandenabeele et al., 2001; Widjaja et al., 2008) |

## 3.2 1D CNN for Raman Spectrum Recognition

One-dimensional CNN is a multivariate method. It is first used for Raman spectroscopy application in two concurrent studies by (Acquarelli et al., 2017; Jinchao Liu et al., 2017). Prior to these two studies, preprocessing of the Raman spectroscopy data such as cosmic ray removal, smoothing, and baseline correction was an essential part of machine learning techniques. Furthermore, the dimensionality of the data is often reduced using principal components analysis (PCA) prior to the application of CNN analysis. By contrast, the 1D CNN model trained on raw spectra significantly outperformed other machine learning methods such as SVM (X. Chen et al., 2019; Jinchao Liu et al., 2017; Malek et al., 2018). (Acquarelli et al., 2017) investigated the usage of one layer CNN model for several publicly available spectroscopic datasets. Their results indicate that CNN is less dependent on preprocessing than advanced conventional techniques such as PLS-LDA. A modified 1D version of LeNets (LeCun et al., 1998) model was tested on a preprocessed dataset of the RRUFF database (Jinchao Liu et al., 2017). Their classification accuracy reported around 88% in comparison with the performance of SVM with accuracy near 81%. The same research group conducted another test on a different dataset of the RRUFF database without any preprocessing, and CNN could achieve significant accuracy of 93.3%. In a similar study (Fan et al., 2019), 1D CNN (named DeepCID) was utilized for the identification of six mixtures and 167 component identification datasets, and they claimed that the true positive (TP) rate of 100% and classification accuracy of 98.8%, respectively. Different hyperparameters, such as the number of convolutional layers and kernel sizes, are assessed using two 1D CNN structures by (Hu et al., 2019). The authors claimed that they could reach the accuracy of 100% for a dataset of mine water using more than two 1D convolutional layers. The same model was utilized for Amino Acids datasets with $R^2$ accuracy of ~0.98 (X. Yan et al., 2020). In another application (human blood discrimination from the blood of an animal), a 1D CNN model with two new layers of denoising and baseline correction layers could reach ~94% accuracy for human and animal blood sample classification problems (Dong et al., 2019).

## 3.3 Recent Progress of 1D CNN for Spectroscopy

Satisfying results of 1D CNN for spectral data analysis (Yang et al., 2019) revealed the potential of using deep learning methods for spectral analysis from the classification of one substance to mixture component identification in various fields of science (Lussier et al., 2020). The majority of recent publications use 1D CNN for different spectral applications. However, few studies highlighted the need for further research to address the lack of enough samples relative to the number of features. One method to increase the number of samples in a database and avoid over-fitting is data augmentation. Augmenting spectral datasets is accomplished by adding various offsets, slopes, and multiplications on the vertical axis (Bjerrum et al., 2017) and small shifts along the horizontal direction. Adding noise, preprocessing data, and constructing new spectra by averaging over several spectra of a sample can be other options of data augmentation.

Another alternative method of increasing performance without overfitting is transfer learning (sometimes called Domain adaptation). This method is useful for the cases that the knowledge (network weights) of a 1D CNN model well-trained on one larger dataset is used for training and testing on a novel smaller dataset with different characteristics. This case is more likely to happen for handheld spectrometers, where we want to use a high-resolution reference standard library captured in the lab for identification of a test spectrum with a different distribution captured by a handheld spectrometer. Domain adaption can be used to enable one CNN model to work well on both domains albeit at the expense of a small accuracy reduction, depends on the dissimilarity features of the source and the target domains. For a typical 1D CNN model, the majority of trainable parameters are in the last layer in the form of fully connected or dense layers. During transfer learning, trained weights from the encoder part of a source network are frozen, and only the last layer of the network is trained again on the target domain. Simple transfer learning on Raman spectra identification has been studied recently by (R. Zhang et al., 2020). The authors used parameters of 1D CNN trained on one database for training and testing on a new database with different characteristics. Results of using domain adaptation are usually more satisfying than training a model from scratch. (R. Zhang et al., 2020) could reach the accuracy of 88.4% and 94% with and without preprocessing the data, respectively. Their source domain data were a subset of spectra from Bio-Rad database, and Raman spectral data in target domain was IDSpec ARCTIC Raman spectrometer. Another application of domain adaptation for Raman spectroscopy has been investigated by (Ho et al., 2019) for the Bacterial database with 99.7% classification accuracy. Spectral mapping can be considered as another approach where high-resolution laboratory spectra are mapped on lower resolution sprctra for unknown material identification applications (Weatherall et al., 2013). Although this technique is useful, all data should be transformed for mapping, followed by several steps of preprocessing. Therefore, this method cannot be a fully-automatic, real-time performance, and end-to-end technique.

One difficulty in the training of deep CNN architectures with a tenth of layers is vanishing gradient problem. As the gradient is back-propagated toward the input layer, repeated multiplication between layers causes the gradient to become infinitely small. As a result, the output of the network saturated to zero, and it is not feasible to train profound CNN models. This difficulty has been addressed by using identity shortcut connections in a new architecture named Residual neural network (ResNet). In this model, by directly bypassing the input information to the output layer, the integrity of the information can be protected. ResNet has been used for the first time for Raman spectral classification task in a few studies (X. Chen et al., 2019; Ho et al., 2019). 1D ResNet model was evaluated on a small Raman spectra-encoded suspension arrays dataset with 15 classes. Compared with a typical 1D CNN without residual blocks, the classification accuracy of the ResNet model is boosted less than 1%, which is not significant (X. Chen et al., 2019).

**Error! Reference source not found.** and Table 5 show a list of publications and some of their databases, respectively, that use 1D CNN in Raman Spectral analysis. As can be seen from the **Error! Reference source not found.**, CNN could achieve better results (more than 90% classification accuracy) in almost studies in comparison to conventional techniques., The result of using raw databases is significantly better than using preprocessed databases in some studies. One reason for this improvement maybe that the preprocessed data tends to retain potentially valuable and sometimes not immediately obvious information in the form of raw spectra which maybe partially lost in the processed dataset. Unfortunately, due to the lake of using data augmentation in some studies and the use of small database in comparison to the network size, there is telltale signs of overfitting. This suggests further research efforts is needed to further improve the 1D CNN for spectral applications.

Evaluation results of using 1D CNNs for other types of vibrational spectroscopy also indicated an outstanding performance over previously proposed machine learning techniques. (Malek et al., 2018) optimized 1D CNN model for near-infrared (NIR) regression problems using a well know heuristic optimization method (Particle Swarm Optimization (Kennedy & Eberhart, 1995)). As a completion of (Malek et al., 2018) work, (Cui & Fearn, 2018) investigates the application of 1D CNN for multivariate regression for NIR calibration. Both studies claim a significant loss of accuracy (~28%) happened after using 1D CNN as a regression model in comparison with previous methods such as PLSR, GPR, and SVR without applying any data preprocessing. Recently, inspired by the Inception Network (Szegedy et al., 2015) using fully convolutional layers, (X. Zhang et al., 2019) proposed a new 1D CNN Inception model (named DeepSpectra) for NIR spectroscopic classification tasks. DeepSpectra outperformed previous methods PCA-ANN, SVR, and PLS while it predicts better results on four different raw datasets than the preprocessed version of those NIR spectra data (X. Zhang et al., 2019). Due to the limited ability of PCA in dimension reduction, such as losing information or weaker performance for nonlinear datasets, Deep auto-encoder (DAE) can be an excellent alternative for PCA methods in spectroscopy with better feature extraction ability (T. Liu et al., 2017). A 1D deep DAE has utilized for NIRS feature extraction in a study by (T. Liu et al., 2017). In a recent study by (Lussier et al., 2020), 1D CNN was used for Surface-Enhanced Raman Scattering (SERS) in Optophysiology application, which is the first use of this method in SERS spectral analysis.

**Table 4. A list of studies used 1D CNN for Raman Spectroscopy. N/A means there are no details provided in the publication.**

| References | CNN Method | Raman Database /# of Samples /# of classes | Data Pre-processing | Data Augmentation | Network Initialization | Classification Accuracy |
|---|---|---|---|---|---|---|
| (Jinchao Liu et al., 2017) | 1D LeNet | RRUFF / 5168 / 1671 | Raw & Cosmic and Baseline correction | H shift & noise | Gaussian | Raw 93.3% & Pre 88.4% |
| (X. Chen et al., 2019) | 1D ResNet | Reporter Molecules /1831 /15 | N/A | - | N/A | 100% |
| (Ho et al., 2019) | 1D ResNet | Bacterial dataset / 2000 / 30 | Baseline correction | Combining Spectra | Pre-trained CNN | 99.7% |
| (Fan et al., 2019) | 1D CNN DeepCID | B&W Tek / 167 / Mix | Raw | N/A | Truncated normal distribution | ~99% |
| (Acquarelli et al., 2017) | 1D CNN CNNVS | Beer & Tablets / 165 / 6 | Raw & Baseline and scattering correction, noise | Combining Spectra | Xavier initialization (Glorot & Bengio, 2010) | Raw ~83.5% & Pre ~91% |

| | | | removal, and scaling | | | |
|---|---|---|---|---|---|---|
| (Hu et al., 2019) | 1D CNN | Mine Water Inrush / 675 / 4 | N/A | N/A | N/A | ~100% |
| (Dong et al., 2019) | 1D CNN | Human and Animal Blood / 326 / 11 | Raw | N/A | N/A | ~95% |
| (Jinchao Liu et al., 2019) | 1D CNN Siamese | RRUFF & CHEMK & UNIPR / 2200 / 230 | Raw & Baseline correction | H shift & noise & combining spectra | Xavier initialization (Glorot & Bengio, 2010) | Raw ~88% & Pre ~92% |
| (R. Zhang et al., 2020) | 1D CNN Transfer Learning | Source Data: Bio-Rad Organics / 5244 / 1685 Target: IDSpec ARCTIC / 216 / 72 | Raw & Baseline correction | H shift & noise | Random | Raw 94% & Pre 88.4% |
| (X. Yan et al., 2020) | 1D CNN | Amino Acids /96 / - | Raw | N/A | Variance scaling | R2 of ~0.98 |
| (Lussier et al., 2020) | 1D CNN | SERS / 1000 / 7 | Baseline correction and noise removal | - | N/A | N/A |
| (W. Lee et al., 2019) | 1D CNN | Extracellular vesicles / 300 / 4 | Raw | Noise | Random | ~93% |
| (H. Yan et al., 2019) | 1D CNN | Tongue squamous cell / - / 2 | Baseline correction and noise removal | N/A | Gaussian | ~95% |

**Table 5. A list of non-personalized datasets has been used for testing 1D CNN techniques.**

| **Name** | **Usage reference** |
|---|---|
| RRUFF (Mineral) | (Lafuente et al., 2016) |
| Azo Pigments | (Vandenabeele et al., 2000) |
| FT Raman Spectra | (Burgio & Clark, 2001) |
| e-VISART | (Castro et al., 2005) |
| Biological molecule dataset | (De Gelder et al., 2007) |
| Explosive compound | (Hwang et al., 2013) |
| Pharmaceutical | (Z. M. Zhang et al., 2014) |
| CHEMK (Chemical) | (Jinchao Liu et al., 2019) |
| UNIPR (Mineral) | (*DiFeSt*, 2020) |
| BioRad (Organics) | (R. Zhang et al., 2020) |

# 4 Conclusion and Outlooks

Assessment of different types of chemicals and biologically hazardous materials in the field requires a fast, portable Raman spectrometer. Rapid identification of substances such as narcotics, explosives, and industrial toxins is a vital necessity for first responders. Modern portable Raman spectrometers have the ability to detect even from barriers. Usage of several portable Raman spectrometers (Jehlička et al., 2017) has been recently investigated on different applications (Izake, 2010). Unlike a laboratory version of spectroscopy, there is limited adoption of artificial intelligence in portable Raman spectroscopy and in portable optical spectrometers. We want to note that many of the AI-based algorithms discussed in this review article are broadly applicable to IR and NIR absorption spectral analysis.

Only a few researchers utilized modern artificial intelligent methods in portable Raman spectrometers (Chandler et al., 2019). The main reason for this slow development of Raman spectrometers is personalized Raman spectrum libraries, where applications are limited to mineral, planetary, medicine, and narcotics. Moreover, expensive Raman spectrometer devices make it difficult for researchers to use handheld systems for their studies. The major part of the cost came from licensing related to matching software as well as built-in Raman libraries, while expenses can be alleviated by using publicly available deep learning techniques and Raman spectral libraries.

Although 1D CNN outperformed other conventional Raman spectral analysis techniques for pure compound identification, there is no extensive development for mixture identification problem. 1D CNN could be trained on databases without the use of preprocessing. It can be achieved automatically and carried out end-to-end. 1D CNN can provide results in real-time. In the cases of the one-dimensional Raman spectra databases, most of the publications used high-end GPU systems for training and testing. Therefore, the usage of CPU power warrants further investigation. We have noticed a number of contradicting reports in several published works. In a few studies, researchers claimed that preprocessing would decrease the accuracy yet their results shows signs of overfitting. In other words, validation steps are not explained, and no ablation studies were carried out in the presented 1D CNN models. One example is the Random Forest method recently proposed for Raman spectroscopy classification (Sevetlidis & Pavlidis, 2019). The authors claimed that the success of deep neural networks comes with complexity and interpretability, and they could reach better results with less complexity.

In conclusion, usage of 1D CNN on Raman spectroscopy indicates a promising alternative to conventional methods where the cumbersome semi-automatic preprocessing phases can be omitted, result in faster and better generalization over different databases. Usage of these techniques for handheld spectrometers to tackle mixtures compound analysis is an open research problem. One success factor for the Deep learning field in computer vision came from publicly available image databases with millions of data. Lack of a similar database for Raman spectral analysis usable for the deep learning field presents the biggest challenge for the researchers in the field.

# 5 References

Acquarelli, J., van Laarhoven, T., Gerretzen, J., Tran, T. N., Buydens, L. M. C., & Marchiori, E. (2017). Convolutional neural networks for vibrational spectroscopic data analysis. *Analytica Chimica Acta*, *954*, 22–31. https://doi.org/10.1016/j.aca.2016.12.010

Arrobas, B. G., Ciaccheri, L., Mencaglia, A. A., Rodriguez-Pulido, F. J., Stinco, C., Gonzalez-Miret, M. L., Heredia, F. J., & Mignani, A. G. (2015). Raman spectroscopy for analyzing anthocyanins of lyophilized blueberries. *2015 IEEE SENSORS - Proceedings*, 1–4. https://doi.org/10.1109/ICSENS.2015.7370224

Baek, S.-J., Park, A., Ahn, Y.-J., & Choo, J. (2015). Baseline correction using asymmetrically reweighted penalized least squares smoothing. *The Analyst*, *140*(1), 250–257. https://doi.org/10.1039/c4an01061b

Bjerrum, E. J., Glahder, M., & Skov, T. (2017). *Data Augmentation of Spectral Data for Convolutional Neural Network (CNN) Based Deep Chemometrics*. 1–10. http://arxiv.org/abs/1710.01927

Burgio, L., & Clark, R. J. H. (2001). Library of FT-Raman spectra of pigments, minerals, pigment media and varnishes, and supplement to existing library of Raman spectra of pigments with visible excitation. In *Spectrochimica Acta - Part A: Molecular and Biomolecular Spectroscopy* (Vol. 57, Issue 7). https://doi.org/10.1016/S1386-1425(00)00495-9

Carey, C., Boucher, T., Mahadevan, S., Bartholomew, P., & Dyar, M. D. (2015). Machine learning tools formineral recognition and classification from Raman spectroscopy. *Journal of Raman Spectroscopy*, *46*(10), 894–903. https://doi.org/10.1002/jrs.4757

Castro, K., Pérez-Alonso, M., Rodríguez-Laso, M. D., Fernández, L. A., & Madariaga, J. M. (2005). On-line FT-Raman and dispersive Raman spectra database of artists' materials (e-VISART database). *Analytical and Bioanalytical Chemistry*, *382*(2), 248–258. https://doi.org/10.1007/s00216-005-3072-0

Chandler, L. L., Huang, B., & Mu, T. (2019). *A smart handheld Raman spectrometer with cloud and AI deep learning algorithm for mixture analysis*. *1098308*(May 2019), 7. https://doi.org/10.1117/12.2519139

Chen, T., Morris, J., & Martin, E. (2007). Gaussian process regression for multivariate spectroscopic calibration. *Chemometrics and Intelligent Laboratory Systems*, *87*(1), 59–71. https://doi.org/10.1016/j.chemolab.2006.09.004

Chen, X., Xie, L., He, Y., Guan, T., Zhou, X., Wang, B., Feng, G., Yu, H., & Ji, Y. (2019). Fast and accurate decoding of Raman spectra-encoded suspension arrays using deep learning. *Analyst*, *144*(14), 4312–4319. https://doi.org/10.1039/c9an00913b

Cui, C., & Fearn, T. (2018). Modern practical convolutional neural networks for multivariate regression: Applications to NIR calibration. *Chemometrics and Intelligent Laboratory Systems*, *182*(July), 9–20. https://doi.org/10.1016/j.chemolab.2018.07.008

De Gelder, J., De Gussem, K., Vandenabeele, P., & Moens, L. (2007). Reference database of Raman spectra of biological molecules. *Journal of Raman Spectroscopy*, *38*(9), 1133–1147. https://doi.org/10.1002/jrs.1734

Deng, L. (2014). Deep Learning: Methods and Applications. *Foundations and Trends® in Signal Processing*, *7*(3–4), 197–387. https://doi.org/10.1561/2000000039

*DiFeSt*. (2020).

Dong, J., Hong, M., Xu, Y., & Zheng, X. (2019). A practical convolutional neural network model for discriminating Raman spectra of human and animal blood. *Journal of Chemometrics*, *33*(11), 1–12. https://doi.org/10.1002/cem.3184

Drucker, H., Wu, D., & Vapnik, V. N. (1999). Support vector machines for spam categorization. *IEEE Transactions on Neural Networks*, *10*(5), 1048–1054. https://doi.org/10.1109/72.788645

Eilers, P. H. C. (2003). A perfect smoother. *Analytical Chemistry*, *75*(14), 3631–3636. https://doi.org/10.1021/ac034173t

Fan, X., Ming, W., Zeng, H., Zhang, Z., & Lu, H. (2019). Deep learning-based component identification for the Raman spectra of mixtures. *Analyst*, *144*(5), 1789–1798. https://doi.org/10.1039/c8an02212g

Fang, Z., Wang, W., Lu, A., Wu, Y., Liu, Y., Yan, C., & Han, C. (2018). Rapid Classification of Honey Varieties by Surface Enhanced Raman Scattering Combining with Deep Learning. *2018 Cross Strait Quad-Regional Radio Science and Wireless Technology Conference, CSQRWC 2018*, 1–3. https://doi.org/10.1109/CSQRWC.2018.8455266

Galvão, R. K. H., Kienitz, K. H., Hadjiloucas, S., Walker, G. C., Bowen, J. W., Soares, S. F. C., & Araújo, M. C. U. (2013). Multivariate analysis of the dielectric response of materials modeled using networks of resistors and capacitors. *IEEE Transactions on Dielectrics and Electrical Insulation*, *20*(3), 995–1008. https://doi.org/10.1109/TDEI.2013.6518970

Glorot, X., & Bengio, Y. (2010). Understanding the difficulty of training deep feedforward neural networks. *Journal of Machine Learning Research*, *9*, 249–256.

Gori, M., & Scarselli, F. (1998). Are multilayer perceptrons adequate for pattern recognition and verification? *IEEE Transactions on Pattern Analysis and Machine Intelligence*, *20*(11), 1121–1132. https://doi.org/10.1109/34.730549

Grotch, S. L. (1970). Matching of mass spectra when peak height is encoded to one bit. *Analytical Chemistry*, *42*(11), 1214–1222. https://doi.org/10.1021/ac60293a007

Ho, C. S., Jean, N., Hogan, C. A., Blackmon, L., Jeffrey, S. S., Holodniy, M., Banaei, N., Saleh, A. A. E., Ermon, S., & Dionne, J. (2019). Rapid identification of pathogenic bacteria using Raman spectroscopy and deep learning. *Nature Communications*, *10*(1). https://doi.org/10.1038/s41467-019-12898-9

Howari, F. M. (2003). Comparison of spectral matching algorithms for identifying natural salt crusts. *Journal of Applied Spectroscopy*, *70*(5), 782–787. https://doi.org/10.1023/B:JAPS.0000008878.45600.9c

Hu, F., Zhou, M., Yan, P., Li, D., Lai, W., Bian, K., & Dai, R. (2019). Identification of mine water inrush using laser-induced fluorescence spectroscopy combined with one-dimensional convolutional neural network. *RSC Advances*, *9*(14), 7673–7679. https://doi.org/10.1039/C9RA00805E

Huang, X., Maier, A., Hornegger, J., & Suykens, J. A. K. (2017). Indefinite kernels in least squares support vector machines and principal component analysis. *Applied and Computational Harmonic Analysis*, *43*(1), 162–172. https://doi.org/10.1016/j.acha.2016.09.001

Hwang, J., Choi, N., Park, A., Park, J. Q., Chung, J. H., Baek, S., Cho, S. G., Baek, S. J., & Choo, J. (2013). Fast and sensitive recognition of various explosive compounds using Raman spectroscopy and principal component analysis. *Journal of Molecular Structure*, *1039*, 130–136. https://doi.org/10.1016/j.molstruc.2013.01.079

Ismail Fawaz, H., Forestier, G., Weber, J., Idoumghar, L., & Muller, P. A. (2019). Deep learning for time series classification: a review. *Data Mining and Knowledge Discovery*, *33*(4), 917–963. https://doi.org/10.1007/s10618-019-00619-1

Izake, E. L. (2010). Forensic and homeland security applications of modern portable Raman spectroscopy. *Forensic Science International*, *202*(1–3), 1–8. https://doi.org/10.1016/j.forsciint.2010.03.020

Jehlička, J., Culka, A., Bersani, D., & Vandenabeele, P. (2017). Comparison of seven portable Raman spectrometers: beryl as a case study. *Journal of Raman Spectroscopy*, *48*(10), 1289–1299. https://doi.org/10.1002/jrs.5214

Jermyn, M., Mercier, J., Aubertin, K., Desroches, J., Urmey, K., Karamchandiani, J., Marple, E., Guiot, M. C., Leblond, F., & Petrecca, K. (2017). Highly accurate detection of cancer in situ with intraoperative, label-free, multimodal optical spectroscopy. *Cancer Research*, *77*(14), 3942–3950. https://doi.org/10.1158/0008-5472.CAN-17-0668

Jouppi, N. P., Young, C., Patil, N., Patterson, D., Agrawal, G., Bajwa, R., Bates, S., Bhatia, S., Boden, N., Borchers, A., Boyle, R., Cantin, P. L., Chao, C., Clark, C., Coriell, J., Daley, M., Dau, M., Dean, J., Gelb, B., … Yoon, D. H. (2017). In-datacenter performance analysis of a tensor processing unit. *Proceedings - International Symposium on Computer Architecture*, *Part F1286*, 1–12. https://doi.org/10.1145/3079856.3080246

Kennedy, J., & Eberhart, R. (1995). 47-Particle Swarm Optimization Proceedings., IEEE International Conference. *Proceedings of ICNN'95 - International Conference on Neural Networks*, *11*(1), 111–117.

Khan, S. S., & Madden, M. G. (2012). New similarity metrics for Raman spectroscopy. *Chemometrics and Intelligent Laboratory Systems*, *114*, 99–108. https://doi.org/10.1016/j.chemolab.2012.03.007

Kingma, D. P., & Ba, J. (2014). Adam: A Method for Stochastic Optimization. *ArXiv Preprint ArXiv:1412.6980*.

Kiranyaz, S., Avci, O., Abdeljaber, O., Ince, T., Gabbouj, M., & Inman, D. J. (2019). *1D Convolutional Neural Networks and Applications: A Survey*. 1–20. http://arxiv.org/abs/1905.03554

Kneen, M. A., & Annegarn, H. J. (1996). Algorithm for fitting XRF, SEM and PIXE X-ray spectra backgrounds. *Nuclear Instruments and Methods in Physics Research Section B: Beam Interactions with Materials and Atoms*, *109–110*, 209–213. https://doi.org/10.1016/0168-583X(95)00908-6

Knock, B. A., Smith, I. C., Wright, D. E., Ridley, R. G., & Kelly, W. (1970). Compound identification by computer matching of low resolution mass spectra. *Analytical Chemistry*, *42*(13), 1516–1520. https://doi.org/10.1021/ac60295a035

Kong, K., Kendall, C., Stone, N., & Notingher, I. (2015). Raman spectroscopy for medical diagnostics - From in-vitro biofluid assays to in-vivo cancer detection. *Advanced Drug Delivery Reviews*, *89*, 121–134. https://doi.org/10.1016/j.addr.2015.03.009

Krizhevsky, A., Sutskever, I., & Hinton, G. E. (2017). ImageNet classification with deep convolutional neural networks. *Communications of the ACM*, *60*(6), 84–90. https://doi.org/10.1145/3065386

Kwiatkowski, A., Gnyba, M., Smulko, J., & Wierzba, P. (2010). Algorithms of chemicals detection using Raman spectra. *Metrology and Measurement Systems*, *17*(4), 549–560. https://doi.org/10.2478/v10178-010-0045-1

Lafuente, B., Downs, R. T., Yang, H., & Stone, N. (2016). The power of databases: The RRUFF project. In *Highlights in Mineralogical Crystallography* (pp. 1–29). Walter de Gruyter GmbH. https://doi.org/10.1515/9783110417104-003

Lecun, Y., Bengio, Y., & Hinton, G. (2015). Deep learning. *Nature*, *521*(7553), 436–444. https://doi.org/10.1038/nature14539

LeCun, Y., Bottou, L., Bengio, Y., & Haffner, P. (1998). Gradient-based learning applied to document recognition. *Proceedings of the IEEE*, *86*(11), 2278–2323. https://doi.org/10.1109/5.726791

Lee, S., Lee, H., & Chung, H. (2013). New discrimination method combining hit quality index based spectral matching and voting. *Analytica Chimica Acta*, *758*, 58–65. https://doi.org/10.1016/j.aca.2012.10.058

Lee, W., Lenferink, A. T. M., Otto, C., & Offerhaus, H. L. (2019). Classifying Raman spectra of extracellular vesicles based on convolutional neural networks for prostate cancer detection. *Journal of Raman Spectroscopy*, *September 2019*, 293–300. https://doi.org/10.1002/jrs.5770

Levina, E., Wagaman, A. S., Callender, A. F., Mandair, G. S., & Morris, M. D. (2007). Estimating the number of pure chemical components in a mixture by maximum likelihood. *Journal of Chemometrics*, *21*(1–2), 24–34. https://doi.org/10.1002/cem.1027

Li, H., Liang, Y., & Xu, Q. (2009). Support vector machines and its applications in chemistry. *Chemometrics and Intelligent Laboratory Systems*, *95*(2), 188–198. https://doi.org/10.1016/j.chemolab.2008.10.007

Li, S., Shan, Y., Zhu, X., Zhang, X., & Ling, G. (2012). Detection of honey adulteration by high fructose corn syrup and maltose syrup using Raman spectroscopy. *Journal of Food Composition and Analysis*, *28*(1), 69–74. https://doi.org/10.1016/j.jfca.2012.07.006

Lieber, C. A., & Mahadevan-Jansen, A. (2003). Automated Method for Subtraction of Fluorescence from Biological Raman Spectra. *Applied Spectroscopy*, *57*(11), 1363–1367. https://doi.org/10.1366/000370203322554518

Liu, Jinchao, Gibson, S. J., Mills, J., & Osadchy, M. (2019). Dynamic spectrum matching with one-shot learning. *Chemometrics and Intelligent Laboratory Systems*, *184*(December 2018), 175–181. https://doi.org/10.1016/j.chemolab.2018.12.005

Liu, Jinchao, Osadchy, M., Ashton, L., Foster, M., Solomon, C. J., & Gibson, S. J. (2017). Deep convolutional neural networks for Raman spectrum recognition: A unified solution. *Analyst*, *142*(21), 4067–4074. https://doi.org/10.1039/c7an01371j

Liu, Juntao, Sun, J., Huang, X., Li, G., & Liu, B. (2015). Goldindec: A novel algorithm for raman spectrum baseline correction. *Applied Spectroscopy*, *69*(7), 834–842. https://doi.org/10.1366/14-07798

Liu, T., Li, Z., Yu, C., & Qin, Y. (2017). NIRS feature extraction based on deep auto-encoder neural network. *Infrared Physics and Technology*, *87*, 124–128. https://doi.org/10.1016/j.infrared.2017.07.015

Liu, W., Wang, Z., Liu, X., Zeng, N., Liu, Y., & Alsaadi, F. E. (2017). A survey of deep neural network architectures and their applications. *Neurocomputing*, *234*(December 2016), 11–26. https://doi.org/10.1016/j.neucom.2016.12.038

Liu, Y., & Yu, Y. (2016). A survey of the baseline correction algorithms for real-time spectroscopy processing. *Real-Time Photonic Measurements, Data Management, and Processing II*, *10026*(November 2016), 100260Q. https://doi.org/10.1117/12.2248177

Lussier, F., Thibault, V., Charron, B., Wallace, G. Q., & Masson, J. F. (2020). Deep learning and artificial intelligence methods for Raman and surface-enhanced Raman scattering. *TrAC - Trends in Analytical Chemistry*, *124*. https://doi.org/10.1016/j.trac.2019.115796

Madden, M. G., & Ryder, A. G. (2003). Machine learning methods for quantitative analysis of Raman spectroscopy data. *Opto-Ireland 2002: Optics and Photonics Technologies and Applications*, *4876*(August 2003), 1130. https://doi.org/10.1117/12.464039

Malek, S., Melgani, F., & Bazi, Y. (2018). One-dimensional convolutional neural networks for spectroscopic signal regression. *Journal of Chemometrics*, *32*(5), 1–17. https://doi.org/10.1002/cem.2977

Nijssen, A., Bakker Schut, T. C., Heule, F., Caspers, P. J., Hayes, D. P., Neumann, M. H. A., & Puppels, G. J. (2002). Discriminating basal cell carcinoma from its surrounding tissue by raman spectroscopy. *Journal of Investigative Dermatology*, *119*(1), 64–69. https://doi.org/10.1046/j.1523-1747.2002.01807.x

Och, F. J., & Ney, H. (2001). Discriminative training and maximum entropy models for statistical machine translation. *Proceedings of the 40th Annual Meeting on Association for Computational Linguistics - ACL '02*, July, 295. https://doi.org/10.3115/1073083.1073133

Park, J. K., Park, A., Yang, S. K., Baek, S. J., Hwang, J., & Choo, J. (2017). Raman spectrum identification based on the correlation score using the weighted segmental hit quality index. *Analyst*, *142*(2), 380–388. https://doi.org/10.1039/c6an02315k

Porro, D., Hdez, N., Talavera, I., Núñez, O., Dago, Á., & Biscay, R. J. (2008). Performance evaluation of relevance vector machines as a nonlinear regression method in real-world chemical spectroscopic data. *Proceedings - International Conference on Pattern Recognition*, 8–11. https://doi.org/10.1109/icpr.2008.4761236

Reynolds, D. A., Quatieri, T. F., & Dunn, R. B. (2000). Speaker verification using adapted Gaussian mixture models. *Digital Signal Processing: A Review Journal*, *10*(1), 19–41. https://doi.org/10.1006/dspr.1999.0361

Rodriguez, J. D., Westenberger, B. J., Buhse, L. F., & Kauffman, J. F. (2011). Quantitative evaluation of the sensitivity of library-based Raman spectral correlation methods. *Analytical Chemistry*, *83*(11), 4061–4067. https://doi.org/10.1021/ac200040b

Ruder, S. (2016). An overview of gradient descent optimization algorithms. *ArXiv Preprint ArXiv:1609.04747*, 1–14.

Schulze, G., Jirasek, A., Yu, M. M. L., Lim, A., Turner, R. F. B., & Blades, M. W. (2005). Investigation of selected baseline removal techniques as candidates for automated implementation. *Applied Spectroscopy*, *59*(5), 545–574. https://doi.org/10.1366/0003702053945985

Sevetlidis, V., & Pavlidis, G. (2019). Effective Raman spectra identification with tree-based methods. *Journal of Cultural Heritage*, *37*, 121–128. https://doi.org/10.1016/j.culher.2018.10.016

Shashilov, V. A., & Lednev, I. K. (2010). Advanced statistical and numerical methods for spectroscopic characterization of protein structural evolution. *Chemical Reviews*, *110*(10), 5692–5713. https://doi.org/10.1021/cr900152h

Stein, S. E., & Scott, D. R. (1994). Optimization and testing of mass spectral library search algorithms for compound identification. *Journal of the American Society for Mass Spectrometry*, *5*(9), 859–866. https://doi.org/10.1016/1044-0305(94)87009-8

Stoltzfus, J. C. (2011). Logistic regression: A brief primer. *Academic Emergency Medicine*, *18*(10), 1099–1104. https://doi.org/10.1111/j.1553-2712.2011.01185.x

Svensson, A., & Schön, T. B. (2017). A flexible state–space model for learning nonlinear dynamical systems. *Automatica*, *80*, 189–199. https://doi.org/10.1016/j.automatica.2017.02.030

Szegedy, C., Wei Liu, Yangqing Jia, Sermanet, P., Reed, S., Anguelov, D., Erhan, D., Vanhoucke, V., & Rabinovich, A. (2015). Going deeper with convolutions. *2015 IEEE Conference on Computer Vision and Pattern Recognition (CVPR)*, *91*(8), 1–9. https://doi.org/10.1109/CVPR.2015.7298594

Vandenabeele, P., Hardy, A., Edwards, H. G. M., & Moens, L. (2001). Evaluation of a principal components-based searching algorithm for Raman spectroscopic identification of organic pigments in 20th century artwork. *Applied Spectroscopy*, *55*(5), 525–533. https://doi.org/10.1366/0003702011952307

Vandenabeele, P., Moens, L., Edwards, H. G. M., & Dams, R. (2000). Raman spectroscopic

database of azo and application to modern art studies. *Journal of Raman Spectroscopy*, *31*(6), 509–517. https://doi.org/10.1002/1097-4555(200006)31:6<509::AID-JRS566>3.0.CO;2-0

Weatherall, J. C., Barber, J., Brauer, C. S., Johnson, T. J., Su, Y. F., Ball, C. D., Smith, B. T., Cox, R., Steinke, R., McDaniel, P., & Wasserzug, L. (2013). Adapting Raman spectra from laboratory spectrometers to portable detection libraries. *Applied Spectroscopy*, *67*(2), 149–157. https://doi.org/10.1366/12-06759

Widjaja, E., Zheng, W., & Huang, Z. (2008). Classification of colonic tissues using near-infrared Raman spectroscopy and support vector machines. *International Journal of Oncology*, *32*(3), 653–662. https://doi.org/10.3892/ijo.32.3.653

Wold, S., Sjöström, M., & Eriksson, L. (2001). PLS-regression: A basic tool of chemometrics. *Chemometrics and Intelligent Laboratory Systems*, *58*(2), 109–130. https://doi.org/10.1016/S0169-7439(01)00155-1

Xu, B., Wang, N., Chen, T., & Li, M. (2015). Empirical Evaluation of Rectified Activations in Convolutional Network. *ArXiv Preprint ArXiv:1505.00853*.

Yan, H., Yu, M., Xia, J., Zhu, L., Zhang, T., & Zhu, Z. (2019). Tongue squamous cell carcinoma discrimination with Raman spectroscopy and convolution neural networks. *Vibrational Spectroscopy*, *103*(1), 102938. https://doi.org/10.1016/j.vibspec.2019.102938

Yan, X., Zhang, S., Fu, H., & Qu, H. (2020). Combining convolutional neural networks and on-line Raman spectroscopy for monitoring the Cornu Caprae Hircus hydrolysis process. *Spectrochimica Acta - Part A: Molecular and Biomolecular Spectroscopy*, *226*, 117589. https://doi.org/10.1016/j.saa.2019.117589

Yang, J., Xu, J., Zhang, X., Wu, C., Lin, T., & Ying, Y. (2019). Deep learning for vibrational spectral analysis: Recent progress and a practical guide. *Analytica Chimica Acta*, *1081*, 6–17. https://doi.org/10.1016/j.aca.2019.06.012

Zhang, R., Xie, H., Cai, S., Hu, Y., Liu, G., Hong, W., & Tian, Z. (2020). Transfer-learning-based Raman spectra identification. *Journal of Raman Spectroscopy*, *51*(1), 176–186. https://doi.org/10.1002/jrs.5750

Zhang, X., Lin, T., Xu, J., Luo, X., & Ying, Y. (2019). DeepSpectra: An end-to-end deep learning approach for quantitative spectral analysis. *Analytica Chimica Acta*, *1058*, 48–57. https://doi.org/10.1016/j.aca.2019.01.002

Zhang, Z. M., Chen, S., & Liang, Y. Z. (2010). Baseline correction using adaptive iteratively reweighted penalized least squares. *Analyst*, *135*(5), 1138–1146. https://doi.org/10.1039/b922045c

Zhang, Z. M., Chen, X. Q., Lu, H. M., Liang, Y. Z., Fan, W., Xu, D., Zhou, J., Ye, F., & Yang, Z. Y. (2014). Mixture analysis using reverse searching and non-negative least squares. *Chemometrics and Intelligent Laboratory Systems*, *137*, 10–20. https://doi.org/10.1016/j.chemolab.2014.06.002

Zheng, S., Jayasumana, S., Romera-Paredes, B., Vineet, V., Su, Z., Du, D., Huang, C., & Torr, P. H. S. (2015). Conditional random fields as recurrent neural networks. *Proceedings of the IEEE International Conference on Computer Vision*, *2015 Inter*, 1529–1537. https://doi.org/10.1109/ICCV.2015.179